\documentstyle[emulateapj5,epsf]{aastex}
\def \farcs{\hbox{$.\!\!^{\prime\prime}$}}

\def\etal{et al.\,}

\lefthead{Hoekstra, Wu \& Udalski}
\righthead{Detecting blends with eclipsing binaries in crowded fields}

\begin{document}

\title{An algorithm to detect blends with eclipsing binaries in planet
transit searches}

\author{Henk Hoekstra\altaffilmark{1,2,3}, Yanqin Wu\altaffilmark{3} \& 
Andrzej Udalski\altaffilmark{4}}
\altaffiltext{1}{Department of Physics and Astronomy, University of Victoria,
Victoria, BC, V8P 5C2, Canada}
\altaffiltext{2}{CITA, University of Toronto, Toronto, Ontario M5S 3H8, Canada}
\altaffiltext{3}{Department of Astronomy \& Astrophysics,
        University of Toronto, 60 St. George Street, Toronto, Ontario
        M5S 3H8, Canada}
\altaffiltext{4}{Warsaw University Observatory, Al. Ujazdowskie 4, 00-478 Warszawa,
Poland}
 
\begin{abstract}

We present an algorithm that can detect blends of bright stars with
fainter, un-associated eclipsing binaries. Such systems contaminate
searches for transiting planets, in particular in crowded fields where
blends are common. Spectroscopic follow-up observations on large
aperture telescopes have been used to reject these blends, but the
results are not always conclusive. Our approach exploits the fact that
a blend with a eclipsing binary changes its shape during eclipse.

We analyze original imaging data from the Optical Gravitational
Lensing Experiment (OGLE), which were used to discover planet transit
candidates. Adopting a technique developed in weak gravitational lensing to
carefully correct for the point spread function which varies both with
time and across the field, we demonstrate that ellipticities can be
measured with great accuracy using an ensemble of images.  Applied to
OGLE-TR-3 and OGLE-TR-56, two of the planetary transit candidates, we
show that both systems are blended with fainter stars, as are most
other stars in the OGLE fields. Moreover, while we do not detect shape
change when TR-56 undergoes transits, TR-3 exhibits a significant
shape change during eclipses. We therefore conclude that TR-3 is
indeed a blend with an eclipsing binary, as has been suggested from
other lines of evidence. The probability that its shape change is
caused by residual systematics is found to be less than $0.6\%$.

Our technique incurs no follow-up cost and requires little human
interaction. As such it could become part of the data pipeline for any
planetary transit search to minimize contamination by blends. We
briefly discuss its relevance for the Kepler mission and for binary
star detection.

\end{abstract}

\keywords{binaries:eclipsing $-$ stars: planetary systems}

\section{Introduction}

Discovering planets outside the solar system is one of the key goals
of modern astronomy. Since the first detection (Mayor \& Queloz 1995)
using the radial velocity technique, we have come to know of the
existence of $\sim 140$ extra-solar planets. While radial velocity
monitoring of nearby stars remains the most successful technique in
this venture, a promising alternative is slowly gaining ground. This
so-call `transit' method focuses on detecting planets that transit
their host stars. It requires continuous observing of a large number
of stars, but can provide independent information concerning planet
characteristics otherwise unobtainable by the radial velocity
technique. A couple dozens planet transit searches are currently
underway (see Horne 2003 for a review). Among these, the Optical
Gravitational Lensing Experiment (OGLE) has announced a large number
of planetary transit candidates (Udalski et al. 2002a; and
follow-ups), and a number of these candidates have been confirmed
spectroscopically (e.g., Konacki et al. 2003a, 2004a, 2004b; Bouchy et
al. 2004; Pont et al. 2004; Torres et al. 2005). These are selected to
resemble close-in gaseous planets (period $\leq 10$ days) transiting
main-sequence stars in the galactic disk. Many of the OGLE candidates
(e.g. OGLE-TR-56, $\sim 1.2$ days) have orbital periods a factor of
$2$ shorter than the closest planets discovered by the radial velocity
technique. The questions arise whether this is a new population of
planets and why they are not seen by the radial velocity method. If
confirmed to be genuine planets, they pose intriguing challenges for
understanding planet formation, migration and survival.

What types of objects can masquerade as planet transits?  The success
of OGLE in detecting planet transits relies partly on the extreme
crowdiness and hence large base numbers in its fields. However, this
advantage also brings on the masqueraders -- a faint eclipsing binary
system can project coincidentally (or in some cases, associate
physically) near a brighter disk star. The deep eclipses and the
ellipsoidal variations from the binary are then diluted by the light
from the brighter star into shallow eclipses and little variations out
of transit, mimicking the signatures of a transiting planet.

Sirko \& Paczynski (2003) carefully studied the light-curves of these
candidates and concluded that on average $\sim 50\%$ of these are
contaminations by eclipsing binaries, with the shorter-period ones
more likely to be so. Spectroscopic follow-up of a large number of
these candidates (Konacki et al. 2003a; Dreizler et al. 2003) also
reached a similar conclusion, though at a much greater observational
expense. Moreover, spectroscopic observations are not always able to
separate the blends from genuine planetary objects as the blended main
star may show little or no velocity variations (see, e.g. Torres \etal
2004). High-quality photometric light-curves can be used to rule out
the blends (Seager \& Mallen-Ornelas 2003), but such data are
difficult to obtain for the crowded OGLE fields. It is also possible
to exclude some blending configurations by comparing the observed
light-curves against synthetic light-curves constructed using model
isochrones (Torres \etal 2004, 2005). This latter technique is more
powerful if the blend and the main star are physical triples and
therefore are likely coeval.

Our aim in this work is to provide an independent new method to
recognize blends. Our method is efficient, assembly-line in style, and
robust. It uses original imaging data and does not require any
follow-up work. As such, this method may be broadly adopted in light
of the fact that OGLE and other transiting searches are likely to
produce an increasing number of planet candidates in the
future. Moreover, our method is more suitable for detecting blends
that are not physically associated (coincidental alignment) and
thereby complements the light-curve method of Torres \etal
(2004,2005).

We propose to use the fact that a blended system, albeit unresolved in
the images, always leaves a tell-tale sign: the shapes of their images
are not round. The magnitude of the ellipticity depends on the angular
separation and the relative brightness between the primary star and
the seconary blend. As we show below, we can measure the shape of a
typical blend in the OGLE fields with great precision. Comparison of
the shape in and out of transit allows us to identify blends with
eclipsing binaries. For instance, a star blended with an eclipsing
binary with an undiluted eclipsing depth of $50\%$ is expected to
exhibit a factor of 2 change in its ellipticity between the two
phases. The actual change in shape may be smaller, though still
detectable, as a typical star in the OGLE fields is multiply blended.

The success of this technique depends critically on how well we
measure the shape of a star, in relation to other stars in the same
image. This is where the only major obstacle in this method arises:
the point-spread-function (PSF) varies across the image due to a
multitude of distortions in the photon pathway. It also varies with
time as the pathway changes and the seeing fluctuates. PSF anisotropy
and seeing change the shapes of the objects and renders raw
measurements of the ellipticity unreliable. A similar problem exists
in weak gravitational lensing, where one has to disentangle the
lensing induced distortions in the shapes of faint galaxies from these
observational effects. Fortunately, the weak lensing community has
studied this problem in great detail and has come up with solutions
which we adapt to the case in hand. We note that the method we
develop here have aspects unique to the stellar problem.

Among the hundreds of transiting systems published by the OGLE-III
team (Udalski et al. 2002a, 2002b, 2002c), we choose to focus our
initial efforts on two candidates, OGLE-TR-3 and OGLE-TR-56. On the
basis of spectroscopic follow-up observations with 8m class
telescopes, these two candidates were identified as likely planetary
candidates since their host stars show little or no velocity
variations (Konacki et al. 2003a; Dreizler et al. 2003). TR-56
undergoes genuine flat-bottom transit and has detectable radial
velocity variations, both consistent with a planet explanation.  For
this candidate, the blending scenario was examined in detail by Torres
et al. (2005) who were able to confirm the planetary nature of this
object using a combined analysis of the light curve and radial
velocity measurements. The interpretation for TR-3, however, is more
open to debate. It shows no significant velocity variations, its light
curve contains hints of a secondary eclipse as well as out-of-eclipse
fluctuations (Sirko \& Paczynski 2003; Konacki \etal 2003a). The
method presented here provides a completely independent assessment of
the identities of these two objects.

We briefly describe the data in \S2.  The shape measurement technique
is described in detail in \S3. In \S4 we provide an extensive test of
our analysis and present the results for the two planet transit
candidates in \S5.
 
\section{Data}

The data we analyze were obtained during the third phase of OGLE (OGLE
III, Udalski et al. 2002a). These were collected using the 1.3m Warsaw
telescope at the Las Campanas Observatory, equipped with the 8k MOSAIC
camera. The field of view of the camera is about 35 by 35 arcminutes,
with a pixel scale of $0\farcs26$/pixel. The observations were done in
the $I$-band, and have exposure times of 120s.

Our analysis does not require the full field, so instead we use small
cuts of 600 by 600 pixels, not necessarily centered on the target
candidate. For TR-3 we have 109 images in-transit and 308 images
out-of-transit, whereas we have 65 and 259 images, respectively, for
TR-56. We retrieved all in-transit images, which results in a broad
range in seeing. To minimize the systematic errors caused by the
seeing correction (see \S 3.2), we have selected out-of-transit images
such that their seeing distribution resembles that of the in-transit
data. This is illustrated in Figure~\ref{seeing_dist}.

\vbox{
\begin{center}
\leavevmode
\hbox{%
\epsfxsize=8.5cm
\epsffile[20 150 560 460]{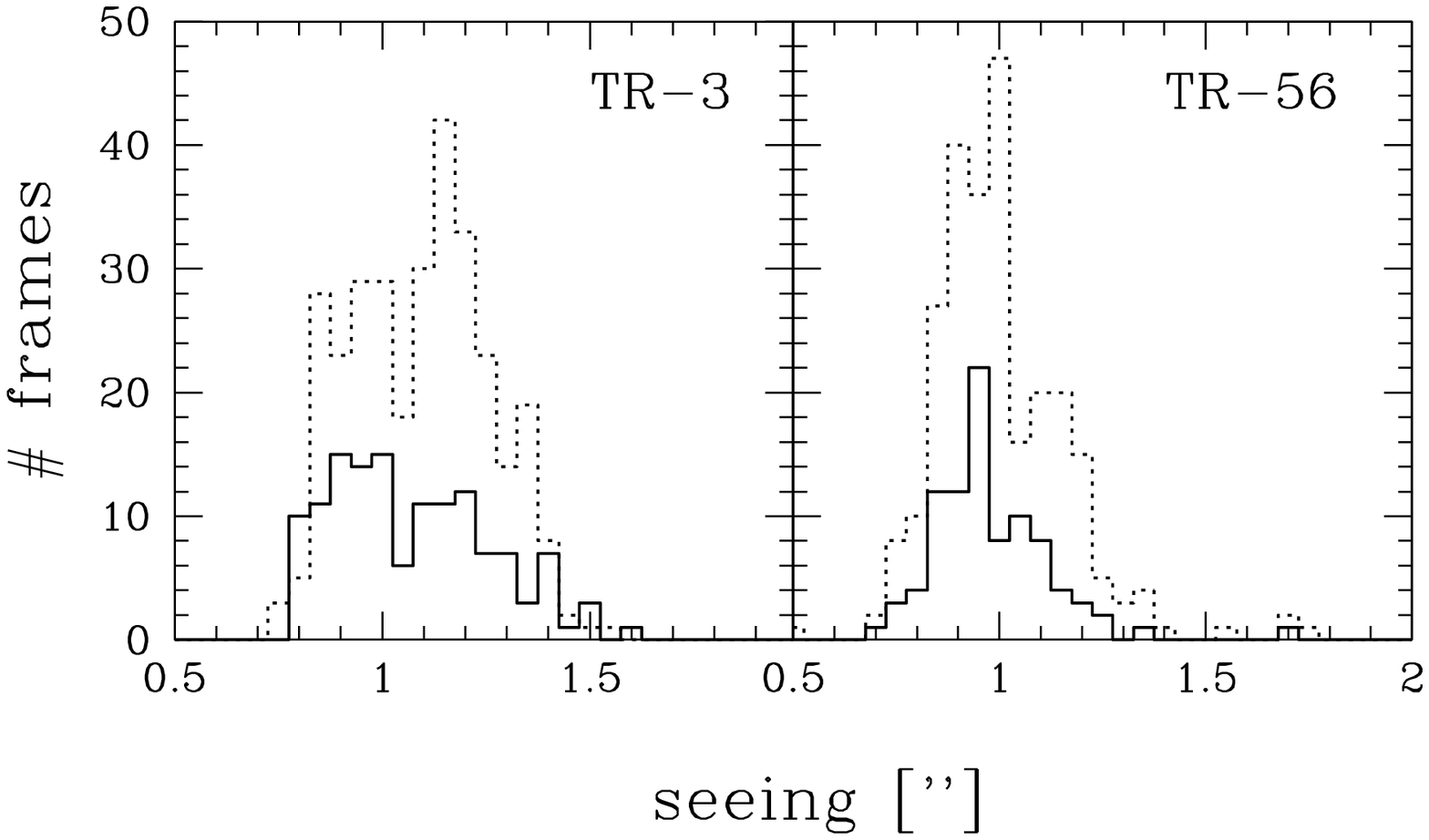}}
\figcaption{\footnotesize Histogram of the seeing distribution of the
images used in the analysis for TR-3 (left) and TR-56 (right). The
solid lines indicate the distribution for the in-transit frame, whereas
the dotted histogram corresponds to the out-of-transit data. The 
seeing distributions of the out-of-transit data were matched to resemble
the in-transit distribution.
\label{seeing_dist}}
\end{center}}

\section{Method}

In this section we discuss the shape measurements, focussing on how to
deal with the variable PSF. The methodology is based on the techniques
developed for weak gravitational lensing applications (e.g., see
Kaiser, Squires \& Broadhurst 1995; Hoekstra et al. 1998), and we
adopt their notation. The correction for the PSF can be split into two
separate steps. The first one is the correction for the anisotropic
part of the PSF, which induces an ellipticity in addition to the
intrinsic ellipticity of the object under investigation. The second
step is the correction for the circularization by the PSF (i.e.,
seeing), which typically lowers the ellipticity. For both steps, we
require a set of comparison stars which can be presumed to be
intrinsically round.

To quantify the shapes, we use the central second moments $I_{ij}$ of
the image fluxes and form the two-component polarisation

\begin{equation}
e_1=\frac{I_{11}-I_{22}}{I_{11}+I_{22}}~{\rm and}~
e_2=\frac{2I_{12}}{I_{11}+I_{22}}.
\end{equation}

Because of photon noise, unweighted second moments cannot be used. Instead
we use a circular Gaussian weight function, with a dispersion $r_s$:

\begin{equation}
I_{ij}=\int d^2{\vec x} f(\vec x)x_i x_j W(|x|,r_s).
\end{equation}

\noindent 

where $x_i$ is the pixel number in the direction of the $i$-axis,
pointing away from the centroid of the object. For the weight function
$W$ we adopt a Gaussian with a dispersion $r_s$.  For the analysis
presented here, the weighted moments are measured from the images
within an aperture with a radius of 6 pixels, and we take $r_s=1.5$
pixels, which is the optimal width for a seeing of 0\farcs9. These
choices suppress the contributions from nearby stars.

\subsection{Anisotropic PSF}

In practice, the PSF will not be isotropic. Instead, the images are
typically concolved with an anisotropic PSF, which induces coherent
ellipticities in the images. In order to recover the true ``shape'' of
the blend, we need to undo the effect of the PSF anisotropy. The
correction scheme we use is based on that developed by Kaiser, Squires
\& Broadhurst (1995), with modifications described in Hoekstra et
al. (1998).

The effect of an anisotropic PSF on the polarisation $e_\alpha$ of an
object is quantified by the ``smear polarisability'' $P^{\rm sm}$,
which measures the response of the polarisation to a convolution with
an anisotropic PSF, and can be estimated for each object from the data
(see Hoekstra et al. 1998 for the correct expressions).

Having measured the polarisations and smear polarisabilities, the
corrected polarisations are given by

\begin{equation}
e_i^{\rm cor}=e_i^{\rm obs}- P^{\rm sm}_{ii} p_i,
\end{equation}

\noindent where $p_i$ is a measure of the PSF anisotropy. It is
measured using a true point source by

\begin{equation}
p_i=e_i^{\rm PSF} / P^{\rm sm,PSF}_{ii},
\end{equation}

\noindent where $e_i^{\rm PSF}$ are the measured ellipticity of the
point source and $P^{\rm sm,PSF}_{ii}$ the diagonal components of its
smear polarisability tensor. Formally, the correction requires the use
of the full two by two tensor, but the off-diagonal terms are
typically small. Examination of the measured values indicates that
they are consistent with noise. We therefore only use the diagonal
terms in the correction for PSF anisotropy.

This correction has been tested extensively in the case of galaxies
convolved with an anisotropic PSF (e.g., Hoekstra et al. 1998; Erben
et al. 2001). For this application, the correction works well, because
galaxies are centrally concentrated, and their shapes are well
characterized by the quadrupole moments.

In the case of two or more nearby point sources the situation is
somewhat different: the shape is not well described by a simple
quadrupole, and higher order moments are expected to contribute to the
polarisation.  To explore this in more detail, we created images that
were convolved with a Moffat (1969) profile and then convolved with a
line (which simulates the PSF anisotropy). A detailed discussion of
this study can be found in the Appendix. Here we summarize the main
conclusion.

The simulations indicate that the correction given by Eqn.~3 is
incomplete and that an additional term proportional to $|p|$ (the
total size of the anisotropy) is needed. This leads to an improved
correction for PSF anisotropy, albeit empirical, given by

\begin{equation}
e_i^{\rm cor}=e_i^{\rm obs}- P^{\rm sm}_{ii} p_i - 
\alpha \sqrt{p_1^2+p_2^2},
\end{equation}

\noindent where the value of $\alpha$ depends on the configuration of
the point sources (separation and flux ratio) and the seeing. We found
that the size of $\alpha$ is proportional to the polarisation of the
object. This is supported by an examination of the residuals in the
shapes of the objects in the OGLE data. 

The fact that $\alpha$ is proportional to the polarisation is not
surprising: when the polarisation is larger, the higher order moments
become more important. However, in the case of OGLE, blends with more
than one source are likely. Consequently, it is difficult to compute
the expected value of $\alpha$. Instead, we determine the value
empirically by fitting a term proportional to $|p|$ to the shape
measurements.

The PSF anisotropy depends on the position of the object on the chip
and it typically varies with time. Fortunately, it is possible to
characterize the spatial variation of the PSF anisotropy with a low
order polynomial model fitted to a subsample of the objects identified
as suitable stars (i.e., the stars should be bright but not
saturated). This works particularly well for the data used here, as we
use relatively small regions around the OGLE transit candidates. For
the analysis here we model the spatial variation by a second order
polynomial. Such a model is derived for each exposure and used to undo
the effect of the PSF anisotropy.

The derivation of the PSF anisotropy model implicitely assumes that
the set of comparison stars are intrinsically round: i.e., the
observed polarisation is solely caused by PSF anisotropy. It is
possible to reject wide separation binaries (or blends) from this set
on the basis of their large ellipticities, but it is more difficult to
reject stars that have a small intrinsic ellipticity because of a
companion. However, so long as the number of comparison stars is
sufficiently large, because their position angles are uncorrelated
with each other and with the PSF anisotropy, we still can obtain an
unbiased model for the PSF anisotropy.

\vbox{
\begin{center}
\leavevmode
\hbox{%
\epsfxsize=8.5cm
\epsffile{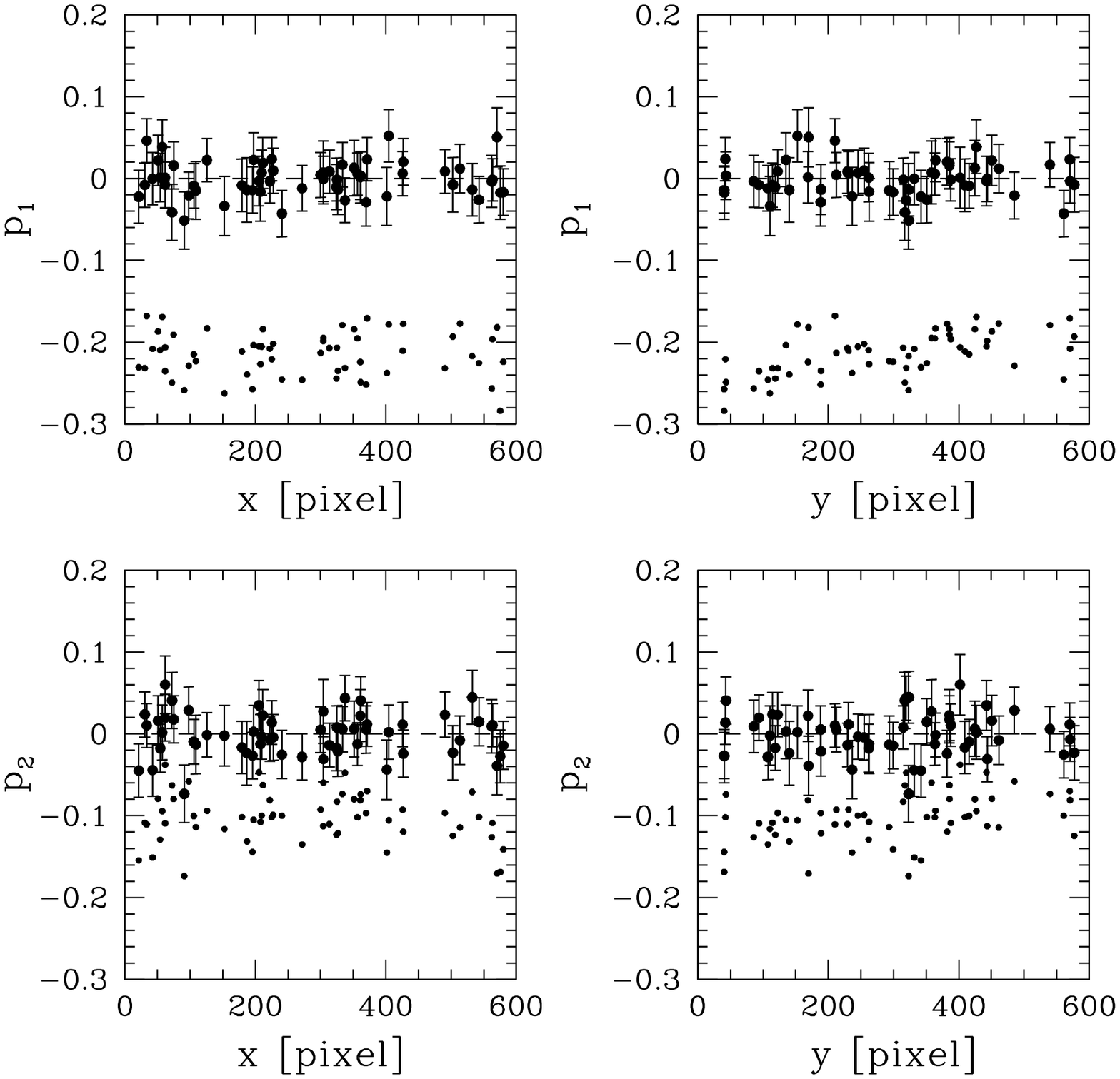}}
\figcaption{\footnotesize 
Residuals in $p_1$ (upper panels) and $p_2$ (lower panels) for the
comparison stars are plotted here (points with error bars) against
position on the chip. They are obtained from the measured values (points
without error bars) after subtracting a second-order polynomial for
the PSF anisotropy. This simple correction leaves no detectable trend
in $p_i$. The scatter may be partially related to the intrinsic
ellipticity arising from blending.
\label{psf_an}}
\end{center}}

In the case of OGLE, severe crowding means most bright stars are
blended with fainter stars. As a result, the noise introduced by the
blends can be substantial. Nevertheless, we can find a set of brighter
stars which are comparatively less affected by blending and provide
good estimates for the PSF anisotropy. The result of this procedure
carried over one frame is presented in Figure~\ref{psf_an}: we
detected significant PSF anisotropy, and found that a second-order
polynomial is sufficient to remove the PSF anisotropy across the whole
image, leaving residual $p_1$ and $p_2$ scattering randomly around the
zero-level.

In the absence of blending and shot noise, the corrected ellipticities
should all be zero (assuming the model used to correct for the PSF
anisotropy is perfect). However, blending gives rise to non-zero
ellipticities for the stars and is partly responsible for the residual
anisotropy in Figure~\ref{psf_an}. We measure this ``intrinsic''
ellipticity of the stars used in the PSF anisotropy correction using
repeated observations taken by the OGLE team (as this procedure
reduces the shot noise). We then subtract the ``intrinsic''
ellipticities from the observed ones and obtain an improved fit. We
found that this iteration had little effect on the results, because of
the random orientations of the blends.

Large values of PSF anisotropy are typically a nuisance, as they imply
larger corrections. However, the large range of PSF anisotropy
(Fig.~\ref{psfan_dist}) exhibited by the OGLE observations is helpful
for the purpose of our paper: it allows us to examine the accuracy of
the correction for PSF anisotropy in more detail, and to understand
the limitation of our algorithm.

\vbox{
\begin{center}
\leavevmode
\hbox{%
\epsfxsize=8.5cm
\epsffile[20 170 550 610]{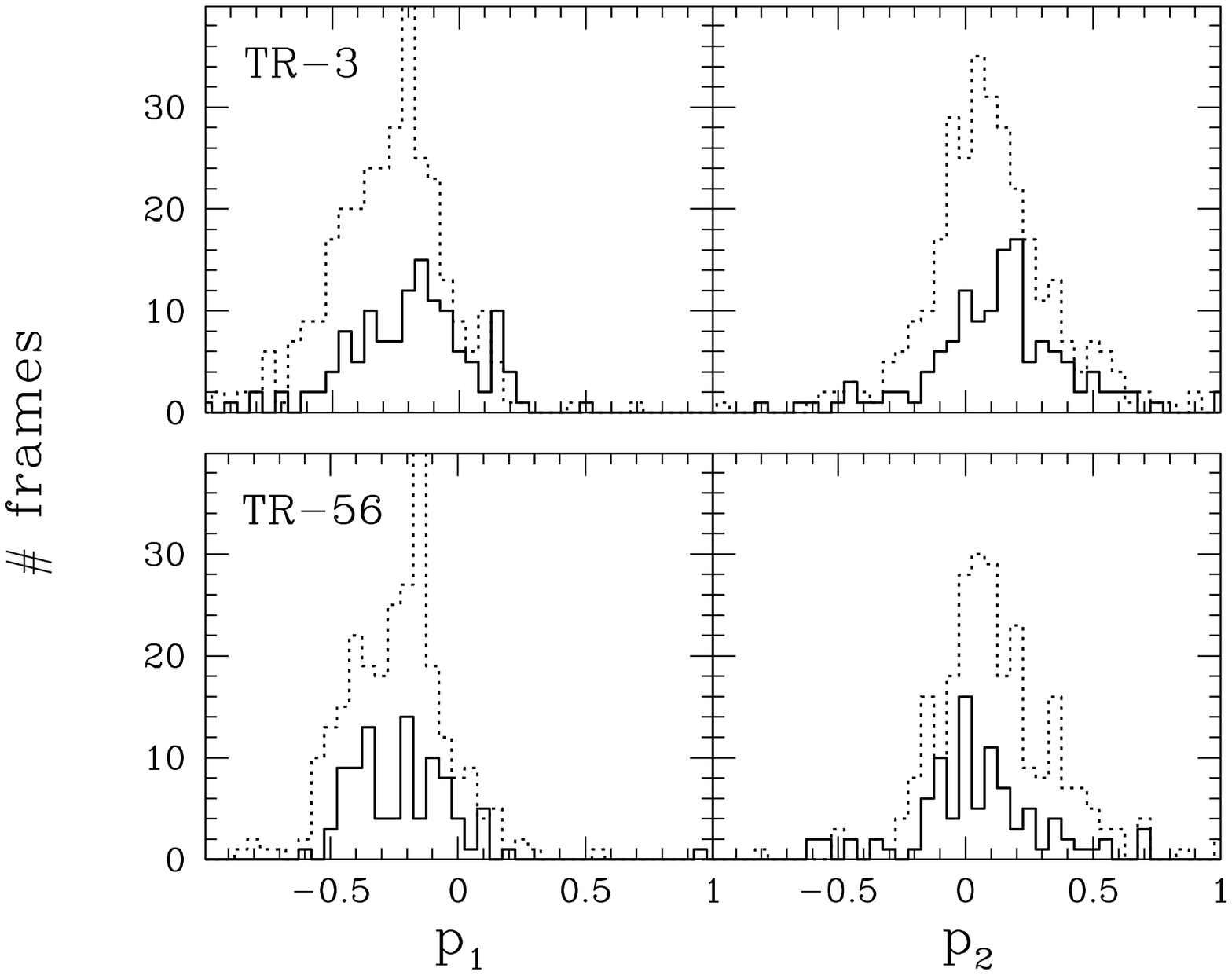}}
\figcaption{\footnotesize Histograms for the measured PSF anisotropy,
$p_1$ and $p_2$ (averaged over all selected stars in an image), for
TR-3 (upper panels) and TR-56 (lower panels) frames. The solid lines
are for the in-transit frames, whereas the dotted lines correspond to
the out-of-transit measurements. The distributions for the in and
out-of-transit frames span similar ranges.
\label{psfan_dist}}
\end{center}}

\subsection{Correction for seeing variation}

The second step in our correction procedure is to account for the
effect of seeing, i.e., the isotropic part of the PSF.  Typically, an
object will appear rounder with increasing seeing. An example is
presented in the left panels of Figure~\ref{seeing_var}, which shows
the ellipticities for one of the stars in TR-3 field as the seeing
varies.  The dependence on seeing can be rather complicated, with some
configurations appearing more eccentric with increasing seeing.

In the simple case of a single blend, one could use simulations to
attempt to determine the seeing dependence. This is not feasible here,
because the target stars are on average blended with 1.5 objects.
Instead, we make use of the fact that the observations span a large
range in seeing (see Fig.~\ref{seeing_dist} and Fig.~\ref{seeing_var})
to remove the seeing dependence empirically. Using the following
model, individually tuned for each star,

\begin{equation}
e_\alpha=a_0 + a_1\cdot seeing+a_2 \cdot seeing^2.
\end{equation}

This second-order fitting is sufficient to remove any visible seeing
dependence for, e.g., the object shown in
Figure~\ref{seeing_var}. From now on, we report the shape measurement
for a fiducial seeing, taken to be 1 arcsecond.

\section{Testing the algorithm}

We selected a sample of a total of 171 stars around the transit
candidate in the two fields. These stars have some range in
brightness and ellipticity. We demonstrate below our capabilities in
removing the effects of PSF anisotropy and seeing.

\vbox{
\begin{center}
\leavevmode
\hbox{%
\epsfxsize=8.5cm
\epsffile[20 160 550 630]{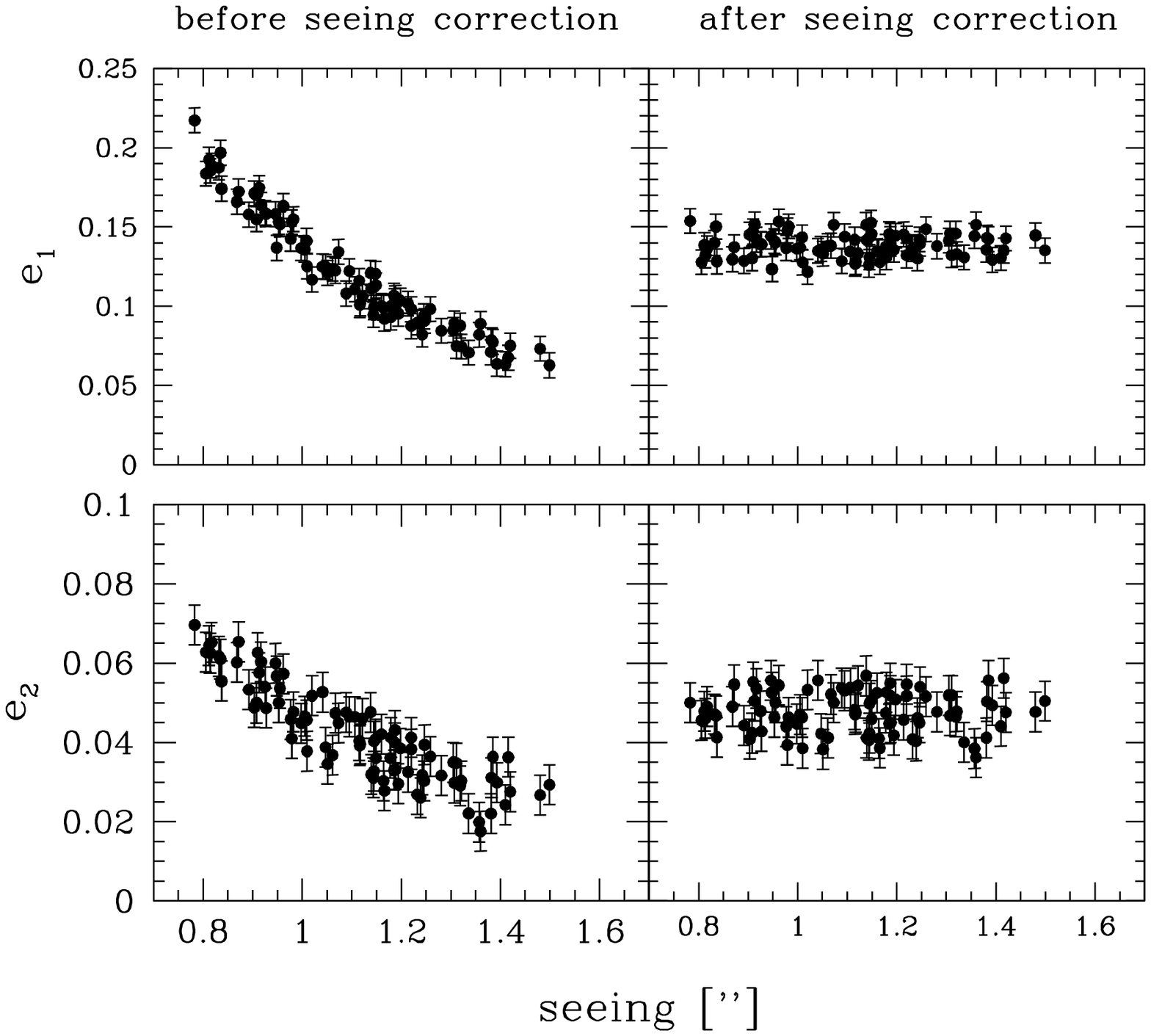}}
\figcaption{\footnotesize {\it Left panels}: variation of the two
components of the polarisation with seeing for an elongated object in
the TR-3 field, after correcting for PSF anisotropy. The polarisation
decreases with increasing seeing, as most stars do. {\it Right
panels}: the value of the polarisation at a fiducial seeing of 1
arcsecond after we account for the variation with seeing using a
second-order polynomial model. Note the different vertical scales for
$e_1$ and $e_2$. The error bars are determined from the scatter in the
polarisations after correcting for seeing.  Although determined
independently, the errors for $e_1$ and $e_2$ are comparable.
\label{seeing_var}}
\end{center}}

\subsection{Correcting for PSF anisotropy}
\label{subsec:PSFa}

To examine the accuracy of the PSF anisotropy correction, we split the
in-transit data into two subsets of similar sizes: one with large PSF
anisotropy ($|p|$) and one with small $|p|$.  The two subsets have a
similar range in seeing. 

In Figure~\ref{psfcor}, we present the differences in ellipticity
measurements for these comparison stars, without correcting for PSF
anisotropy (upper panels), after using equation~(3) to correct for the
anisotropy (middle panels), and after using equation~(5) to correct
for the anisotropy (lower panels).\footnote{This additional correction
is tiny for objects with small ellipticities.} This experiment
convinced us that we can remove PSF anisotropy successfully from our
data. The reduced $\chi^2$ for the results in the lower panels of
Figure~\ref{psfcor} is close to unity, indicating that the estimated
errors are a fair estimate of the statistical uncertainty in the
measurements.

In producing Figure~\ref{psfcor}, we have applied a seeing correction
that is based on the combined in-transit data, minimizing the
systematics caused by the latter correction.  As mentioned above, the
seeing ranges are similar for both samples, so PSF anisotropy is the
only systematic relevant for comparison. Upper panels in
Figure~\ref{shape_zoom} expand the view from the lower panels of
Figure~\ref{psfcor} for the small ellipticity objects, while lower
panels in Figure~\ref{shape_zoom} shows results from the same
procedure using out-of-transit data. For some of the brighter objects,
the achieved error bars are as small as $\sim 1.2\times 10^{-4}$.
This capability to measure shapes accurately brings about another
potential application for the algorithm described here: finding binary
stars that are too close, or are too different in fluxes, to be
resolved (also see \S6).

\vbox{
\begin{center}
\leavevmode
\hbox{%
\epsfxsize=8.5cm
\epsffile[20 180 560 660]{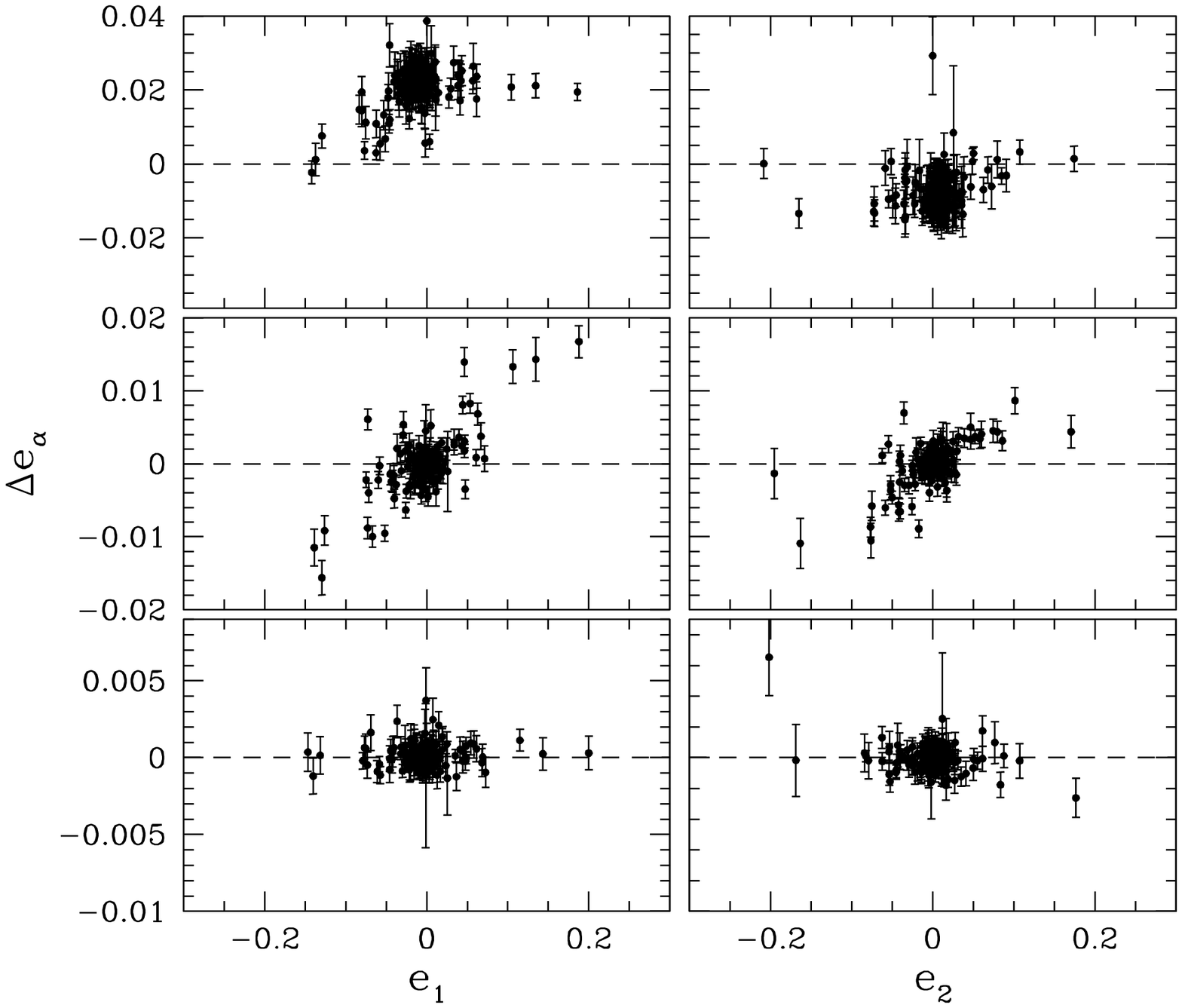}}
\figcaption{\footnotesize {\it Upper panels}: the difference in $e_1$
(left) and $e_2$ (right) between the in-transit data with large PSF
anisotropy and the in-transit data with small anisotropy, before
correction for PSF anisotropy. The offsets from zero simply reflects
the effects of the PSF anisotropy.  {\it Middle panels}: the
difference between the two samples when Eqn.~3 is used to correct for
PSF anisotropy. A clear trend $\propto e_\alpha$ can be
discerned. Note that the vertical scale has been decreased by a factor
two.  {\it Lower panels}: the results when the empirical correction
given by Eqn.~5 is applied for the PSF anisotropy correction. No trend
with the shape of the object is visible. The vertical scale has been
decreased by another factor of two.  The error bars have been
determined from a bootstrap resampling of the data.
\label{psfcor}}
\end{center}}

\vbox{
\begin{center}
\leavevmode
\hbox{%
\epsfxsize=8.5cm
\epsffile[20 170 580 700]{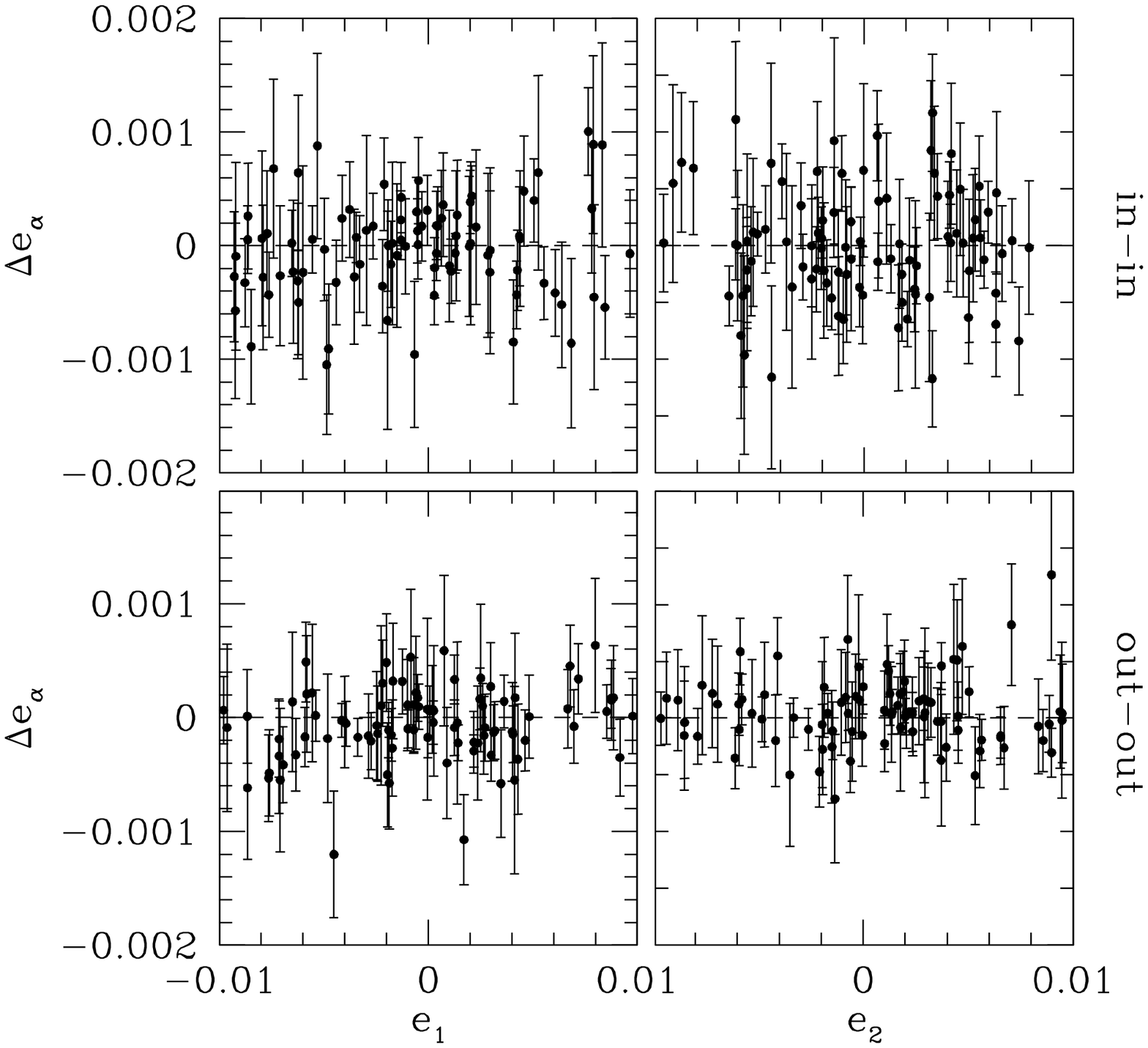}}
\figcaption{\footnotesize {\it Upper panels}: differences in $e_1$
(left) and $e_2$ (right) between the sample with large and small PSF
anisotropy for the in-transit observations, limited to the region of
small ellipticity. {\it Lower panels}: the same differences, but now
between two samples taken from the out-of-transit observations. 
\label{shape_zoom}}
\end{center}}

\subsection{Correcting for Seeing}
\label{subsec:corrects}

We now examine the reliability of the correction for seeing. If this
is successful, we will be able to accurately measure the in and
out-of-transit ellipticity changes in planet candidates, and constrain
the blending scenario.

As shown in Figure~\ref{seeing_dist}, the seeing distributions for the
in and out-of-transit data were chosen to be similar, i.e. we selected
out-of-transit images such that the two distributions match. This
approach minimizes the sensitivity of our results to systematic errors
caused by the adopted seeing correction. We fit equation~(6) to the in
and out-of-transit measurements separately. The resulting differences
between the in and out-of-transit data are presented in
Figure~\ref{difference}. Panels~a and~b show the differences in $e_1$
and and $e_2$ respectively. 

The lower panels in Figure~\ref{difference} show histograms of the
differences in units of the estimated measurement uncertainty. For a
normal distribution, this should be a Gaussian with a dispersion of 2,
which is indicated by the solid smooth line. This Gaussian provides a
fair match to the observed scatter, but the data show more outliers
than what would be expected from a normal distribution. For $\Delta
e_1$ we obtain a reduced $\chi_{\rm red}^2$=1.31 and for $\Delta e_2$
we find a similar value of $\chi_{\rm red}^2$=1.37, larger than the
expected value around unity. These values reduce to 1.14 and 1.16
respectively when we reject objects that are more than $3\sigma$ away
from zero.

The bootstrap analysis provides an estimate of the random error but
not of the systematic error. The results presented in
Figure~\ref{difference} suggest that the estimated errors are
typically correct, but in a few cases, residual systematic errors are
still present in the data leading to the excess of outlyers.
Nevertheless, the results presented in Figure~\ref{difference} do
suggest that for most objects we can measure the difference in shapes
between the in and out-of-transit data accurately. We have attempted
to identify what is causing some of the outlyers, but have not been
able to find an obvious way to improve the measurements. We suspect
that it might be due to imperfections in the correction for PSF
anisotropy.  We also note that some of the objects are not present in
all exposures (as they lie too close to the edge), which might lead to
differences in the actual seeing distributions, which in turn can lead
to systematic errors in the shape differences.

The size of the final error bar as used in Figure~\ref{difference}
depends on the number of frames used as well as on the apparent
magnitude of the object: the shape measurements in a single frame will
be noisier for fainter stars. This is demonstrated in
Figure~\ref{error}, which shows the error in $e_1$ and $e_2$ as a
function of apparent magnitude for the out-of-transit shape
measurements.\footnote{Although the errors in both components of the
polarisation are derived separately, they typically are comparable.}
As expected, the errors increase with magnitude. This is more clearly
seen for ``rounder'' objects, which are affected less by the
correction for seeing and the last term in equation~5 for the PSF
anisotropy correction.

We also computed the smallest possible error bar as a function of
apparent magnitude using simulated images. In these images, which have
the same noise properties as the OGLE data, we measured the scatter in
the shape of a point source. In this case, the error is solely due to
Poisson noise. The result is given by the dashed curve in
Figure~\ref{error}. The actual error bars are larger, because of the
uncertainties introduced by the empirical corrections for PSF
anisotropy and seeing. Finally, Figure~\ref{error} also demonstrates
that the accuracy with which one can measure shapes is excellent: the
typical uncertainty for a star with $m_I=14$ is $\sim 3\times
10^{-4}$.

\vbox{
\begin{center}
\leavevmode
\hbox{%
\epsfxsize=8.5cm
\epsffile[20 170 580 700]{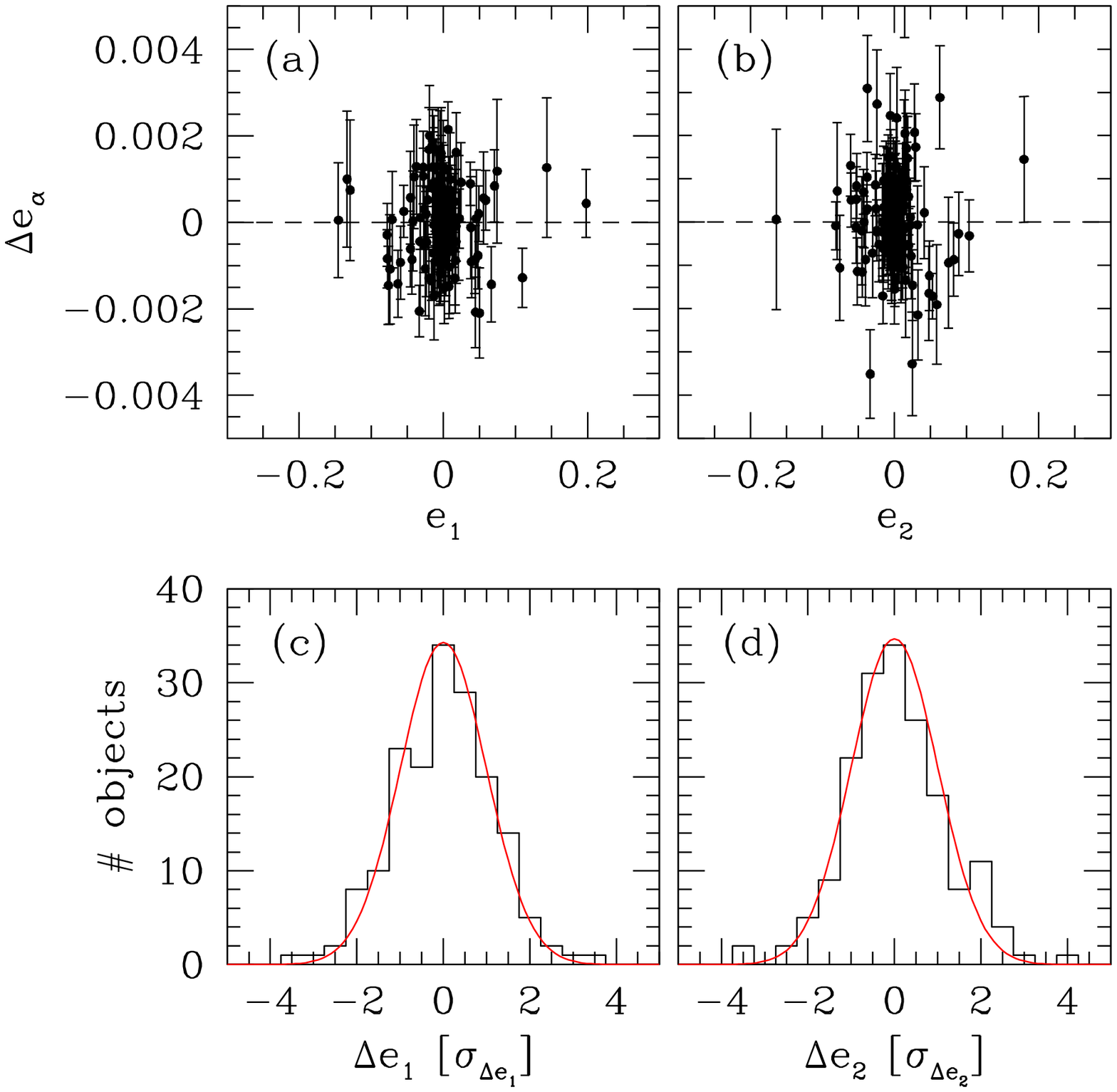}}
\figcaption{\footnotesize {\it panel a}: difference $\Delta e_1$
between the out-of-transit and in-transit measurement of the
polarisation for various stars in the TR-3 and TR-56 fields (note that the
transit candidates are not included in this histogram).  {\it
panel b}: similar to panel~a, but for $\Delta e_2$. For both
components the results suggest that the shape difference can be
measured with great accuracy. {\it panel c}: histogram of the number
of objects as a function of the ratio between $\Delta e_1$ and the
corresponding measurement error. {\it panel d}: the same, but for
$\Delta e_2$. If the errors follow a normal distribution, this should
resemble a Gaussian with a dispersion of 2, which is indicated by the
smooth curve. The agreement with a normal distribution is fair, but
a larger than expected number of outlyers is found, suggestive of
residual systematics. As before, the error bars were obtained from
a bootstrap resampling of the data.
\label{difference}}
\end{center}}

\section{Application to transit candidates}

The results presented in the previous section demonstrates our ability
to accurately measure the shapes of objects in the OGLE fields. In
this section we present results for the two OGLE planet transit
candidates. Table~\ref{tab_shapes}a lists the final polarisations for
TR-3 and TR-56 at a fiducial seeing of 1 arcsecond measured from the
out-of-transit images. We detect a significant polarisation for both
transit candidates, thus implying that they are both blended with
other sources. In fact, most stars studied in the crowded OGLE fields
show evidence of blending (or even multiple blending). Within a circle
of $1''$ radius, an average star is surrounded by $0.6$ companions,
with a mean flux ratio of $4\%$ and a mean sepration of $0.7''$.

The resulting average ellipticity of the blend depends mainly on the
brightness of the primary star: the brighter the star, the smaller the
ellipticity. Figure~\ref{edist} shows the distribution of $e_1$ and
$e_2$ for the analysed stars in the fields of TR-3 and TR-56
(indicated by the crosses). The two transit candidates are indicated
by the open circles, with TR-3 being the point on the left. Although
the distribution is peaked towards round objects, the observed
ellipticities for the transit candidates are by no means anomalous.

\vbox{
\begin{center}
\leavevmode
\hbox{%
\epsfxsize=8.5cm
\epsffile{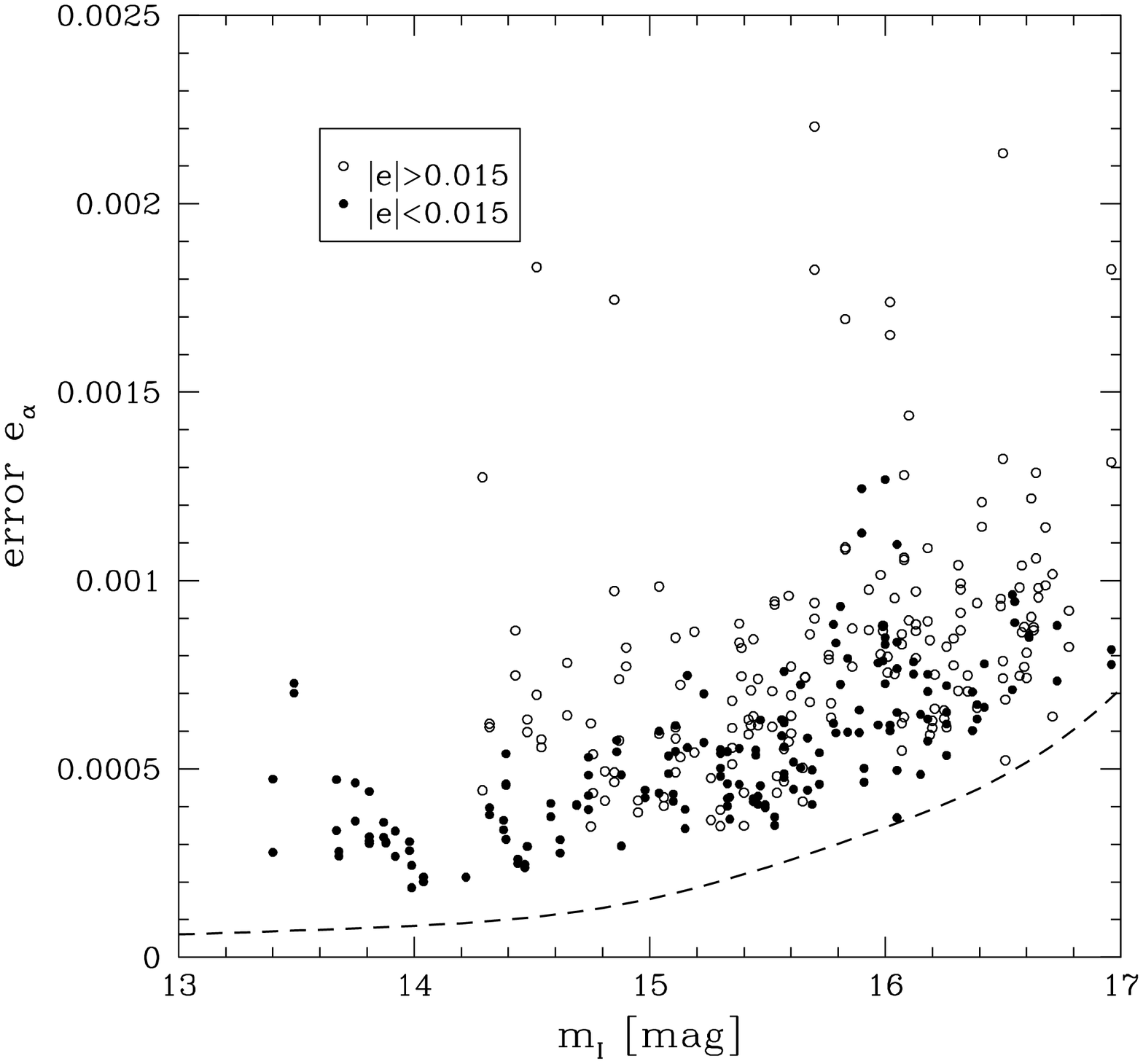}}
\figcaption{\footnotesize Derived errors in $e_1$ and $e_2$ as a
function of apparent magnitude of the object. The solid points
correspond to objects with $|e|<0.015$, i.e., ``rounder'' objects,
whereas the open circles indicate the errors for objects with
$|e|>0.015$. The error bars are larger for more elongated objects
because of the larger uncertainty introduced by the empirical
corrections for PSF anisotropy and seeing. The dashed line indicates
the theoretical minimum value of the error bar as a function of
magnitude. This curve is computed from simulated images of a point
source, with noise properties corresponding the OGLE observations.
\label{error}}
\end{center}}

To test whether the 'planet-like' transit is caused by a blended
eclipsing binary, we also list in Table~\ref{tab_shapes}b differences
in shape between the out and in transit data for these two stars: a
significant change in shape would confirm that the blend is an
eclipsing binary.

\vbox{
\begin{center}
\tabcaption{\footnotesize Results for TR-3 and TR-56, corrected
for PSF anisotropy for a fiducial seeing of 1 arcsecond. \label{tab_shapes}}
\begin{tabular}{llcc}
\hline
\hline
(a) &       & $e_1$ & $e_2$ \\
\hline
    & TR-3  & $-0.0234\pm0.0005$ & $-0.0052\pm0.0005$ \\
    & TR-56 & $ 0.0101\pm0.0005$ & $-0.0005\pm0.0005$ \\
\hline
\hline
(b) &       & $e_1(out)-e_1(in)$  & $e_2(out)-e_2(in)$ \\
\hline
    & TR-3  & $-0.0020\pm0.0005$  & $-0.0007\pm0.0005$ \\
    &TR-56  & $-0.0001\pm0.0006$   & $0.0000\pm0.0006$\\
\hline
\hline
\end{tabular}
\end{center}}

We detect no change in shape in TR-56, suggesting that the observed
transit is a genuine planetary transit, in line with evidences from
radial velocity, line-curve analysis and isochrone fitting (e.g.,
Torres et al. 2005).

In TR-3, however, we do observe a change in shape: the ellipticity
in-transit is larger. The errors inferred from the bootstrap analysis
suggest a significance of $4.2\sigma$. However, the results presented
in Section 4.2 and Figure~\ref{difference} indicate that the
distribution of errors is not exactly Gaussian, but has tails.  We
therefore need to account for the possibility that the change in shape
is caused by residual systematics. To this end, a more conservative
estimate of the significance of the change in shape for TR3 can be
obtained by considering the fraction of studied objects that show a
difference at least as large as TR-3.

\vbox{
\begin{center}
\leavevmode
\hbox{%
\epsfxsize=8.5cm
\epsffile{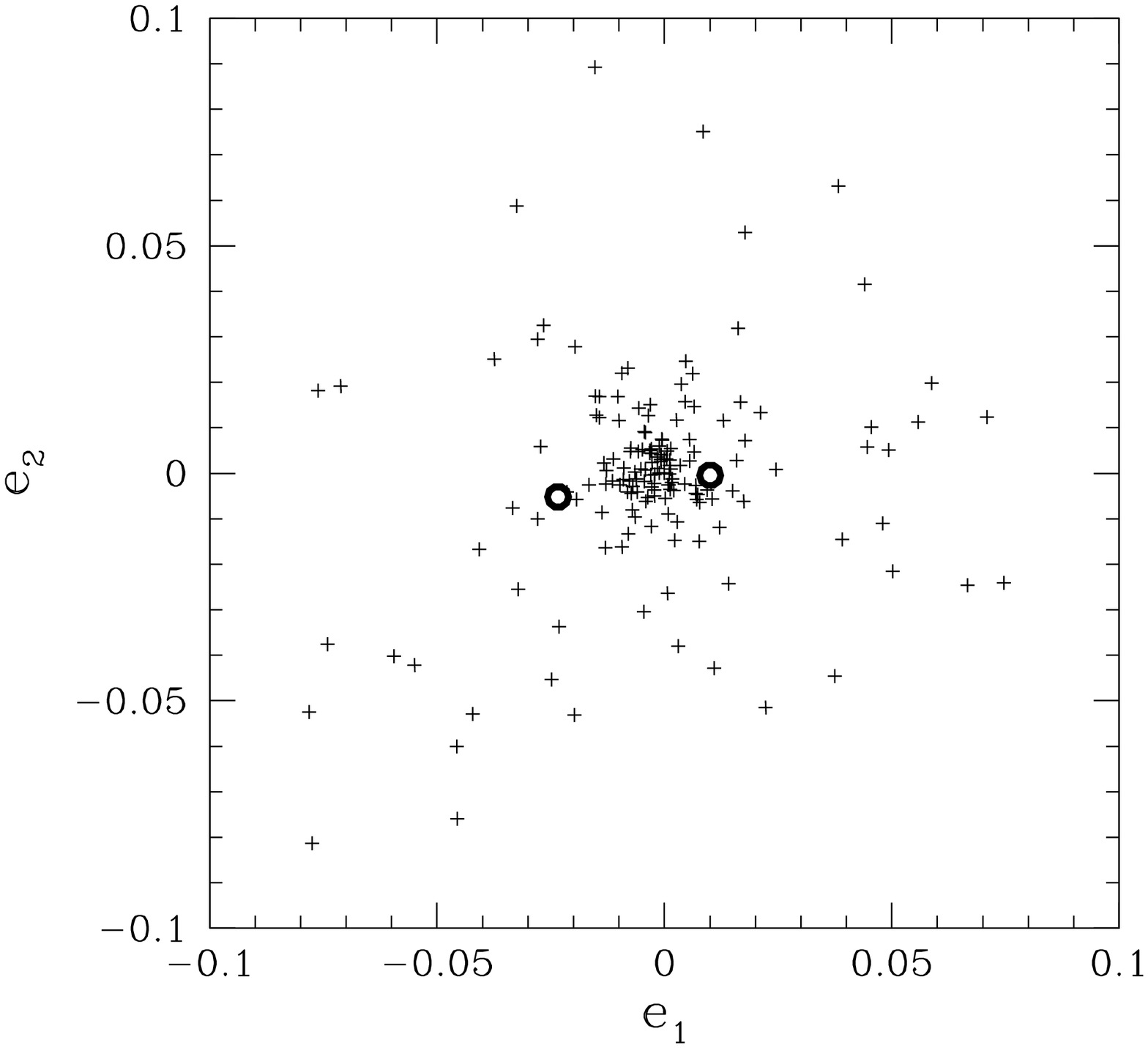}}
\figcaption{\footnotesize The crosses indicate the values of $e_1$ and
$e_2$ for the additional stars that have been analysed in the
out-of-transit images of TR-3 and TR-56. The measurement errors on the
shapes are negligible and have not been displayed. The open circles
indicate the results for TR-3 (left) and TR-56 (right). The
distribution of points is peaked towards round objects, but few
objects have truly no detectable ellipticity. The observed
ellipticities for the transit candidates are by no means anomalous.
\label{edist}}
\end{center}}

Of the objects in the fields of TR-3 and TR-56, accurate shapes could
be determined for 171 of them. None of these objects show an
ellipticity change as large as TR-3, and we can only derive a lower
limit to the probability for the observed shape change in TR-3 to be
caused by systematic effects: the probability is less than $1/171\sim
0.6\%$. This is larger than the probability of a $4-\sigma$ event
($0.03\%$) but still sufficiently small for us to conclude that it is
very likely that TR-3 is indeed a blend with an eclipsing binary
system.

If TR-3 is only singly blended, its polarisation should decrease by a
factor of 2 during eclipse, an effect that should be emminently
detectable. However, we find that its ellipticity increases by $\sim
10\%$ during eclipse.  This can be explained if TR-3 is multiply
blended. In fact, we have also measured TR-3's polarization
(out-of-transit) using different weight functions ($r_s$ in
eq. [2]). It varies with $r_s$ differently than a singly-blended
object would, suggesting that it is indeed multiply blended. In the
case of multiple blending, provided that the primary star is much
brighter than the blending stars, the resulting ellipticity is given
by
\begin{equation}
e_i^{\rm obs}\approx \sum_{j=1}^N e_{i,j} 
\end{equation}
where $N$ is the number of blends, and $e_{i,j}$ is the contribution
from each blend. Consequently, if two blends have opposite signs for
$e_i$, the polarisation can actually increase during an eclipse. 

If we assume that the observed eclipse is caused by a blend with an
eclipsing binary, with a eclipse depth $50\%$ (i.e., a full eclipse of
an equal mass system), we can place limits on the configuration of the
blend. To do so, we note that the depth of the observed transit is
$2\%$, which implies that the flux of the presumed binary contributes
$4\%$ to the total flux. Under these assumptions, the change in
ellipticity indicates that the binary is located $0.38\pm 0.06$ 
arcseconds from the brighter star. If the eclipse depth is reduced
to 25\%, the presumed binary contributes 8\% of the flux instead,
and the separation decreases to $0.26\pm0.04$ arseconds.
These numbers are below the resolution limit of the photometry (by
analysing centroid shift in and out of eclipse, $\sim 1''$), and the
blend is likely a background source (as opposed to a physical triple
with the main star).

Interestingly, by examining the light-curve in detail, Konacki \etal
(2003b) have also come to a similar conclusion that TR-3 is likely a
blend of a background eclipsing binary with a forground bright
star. Our result here confirms their suggestion and predicts the
position of the blend. The sepration from the main star is small but
should be detectable by HST observations.

\section{Conclusions}

We have presented an algorithm that can detect blends of bright stars
with fainter eclipsing binaries. Such systems contaminate searches for
transiting planets, in particular in crowded fields where blends are
common. This technique provides a cheap way to find such blends, thus
minimizing the amount of time required on large aperture telescope for
spectroscopic follow-up of planet candidates.

We have demonstrated the accuracy with which shapes can be measured
using imaging data from the Optical Gravitational Lensing Experiment
(OGLE).  Our method requires a careful correction of the point spread
function which varies both with time and across the field. To this end
we have adopted a method developed in weak gravitational lensing with
modifications necessary for this particular application.

We have tested the correction for PSF anisotropy in great detail,
using a sample of 171 stars surrounding the two planet transit
candidates studied here. Comparison of samples with large and small
PSF anisotropy indicates that this correction can be applied with
great accuracy. For a star with an apparent magnitude $m_I=14$, we
obtain a $1\sigma$ uncertainty of $\sim 3\times 10^{-4}$ in the
polarisation.

Applied to OGLE-TR-3 and OGLE-TR-56, two of the planetary candidates,
we show that both systems are indeed blended with fainter stars, as
are most other stars in the OGLE fields. In the case of TR-56 we do
not detect a change in shape in and out of transit, consistent with it
indeed being a genuine planetary object. For TR-3 we observe a
significant change in shape. If we adopt the error bars from the
bootstrap analysis, the significance is $4.2\sigma$. However, the
distribution of errors is not precisely Gaussian, but has tails. A
more conservative estimate of the significance, estimated from the
observed distribution of shape differences, provides an upper limit of
$0.006$ to the probability that the observed change is caused by
residual systematics. Our results favour the scenario where TR-3 is
caused by a blend with a background eclipsing binary, in line with
evidences from other studies.

A number of studies have appeared since the OGLE announcement of
transit candidates, mostly aiming at distinguishing blends from
genuine planets. In contrast to some of these studies which carry out
follow-up spectroscopy using large telescopes, our approach uses
original imaging data and is a value-added application. Moreover,
unlike studies which perform detailed light-curve fitting or isochrone
stellar model fitting, our method is assembly-line in style and can be
applied to a large number of transit candidates without too much human
interaction. Lastly, our technique is especially suited to finding
blends that are not physically associated with the bright
star,\footnote{These blends cause larger ellipticity.} and is
therefore complementary to the isochrone fitting technique which is
more powerful for the physical triple case. 

Given the efficiency in dealing with a large number of objects without
requiring additional data, the shape method may also be useful for
other planetary transit searches, in particular the NASA Kepler
mission. This transit mission aims to detect $\sim 10^3$ giant inner
planets and $\sim 10^2$ terrestrial planets.  Recently the target
survey area has been moved to a higher galactic latittude to reduce
the confusion by blends with eclipsing binaries.  A quick examination
of the USNO-B catalogue (Monet \etal 2003) in this new field suggests
that the stellar density is $\sim 8$ times less dense than that in the
OGLE field, with a similar number distribution in stellar
magnitudes. However, stars in Kepler have a PSF of $6''$ radius, we
therefore expect each bright star ($m_v < 14$) to have $\sim 2.4$
companions within the PSF envelope, compared to $0.6$ ($< 1''$) in the
OGLE case. The probability of blending with an eclipsing binary is
likely enhanced by a similar ratio. More study is necessary to
determine the false-positive rate due to blending in Kepler, armed
with the experience from OGLE. Nevertheless, we expect that our shape
technique can be readily applied to this mission.

The achieved accuracy in measuring the shape of stars also bodes well
for another potential application of our algorithm: finding binary
stars that are too close to be resolved, yet too far apart for radial
velocity studies. By detecting small deviations from circularity, we
should be able to discover intermediate separation binaries ($\sim 10
- 1000 AU$) with flux ratio as low as $1\%$, within a large volumn of
our galaxy. This will not only complement existing binary searches,
but its high efficiency may also disclose binary population with an
unprecedented rate such as to enable new and meaningful statistical
studies. In a subsequent paper we will investigate this application in
more detail, and apply it to wide field imaging data from the EXPLORE
project (Mall{\'e}n-Ornelas et al. 20003; Yee et al. 2003), which were
obtained with the aim of finding transiting planets.

\acknowledgments AU acknowledges support from the Polish KBN grant
2P03D02124 and the grant ``Subsydium Profesorskie'' of the Foundation
for Polish Science.

\appendix

\begin{figure*}[!t]
\begin{center}
\leavevmode
\hbox{%
\epsfxsize=7.5cm
\epsffile{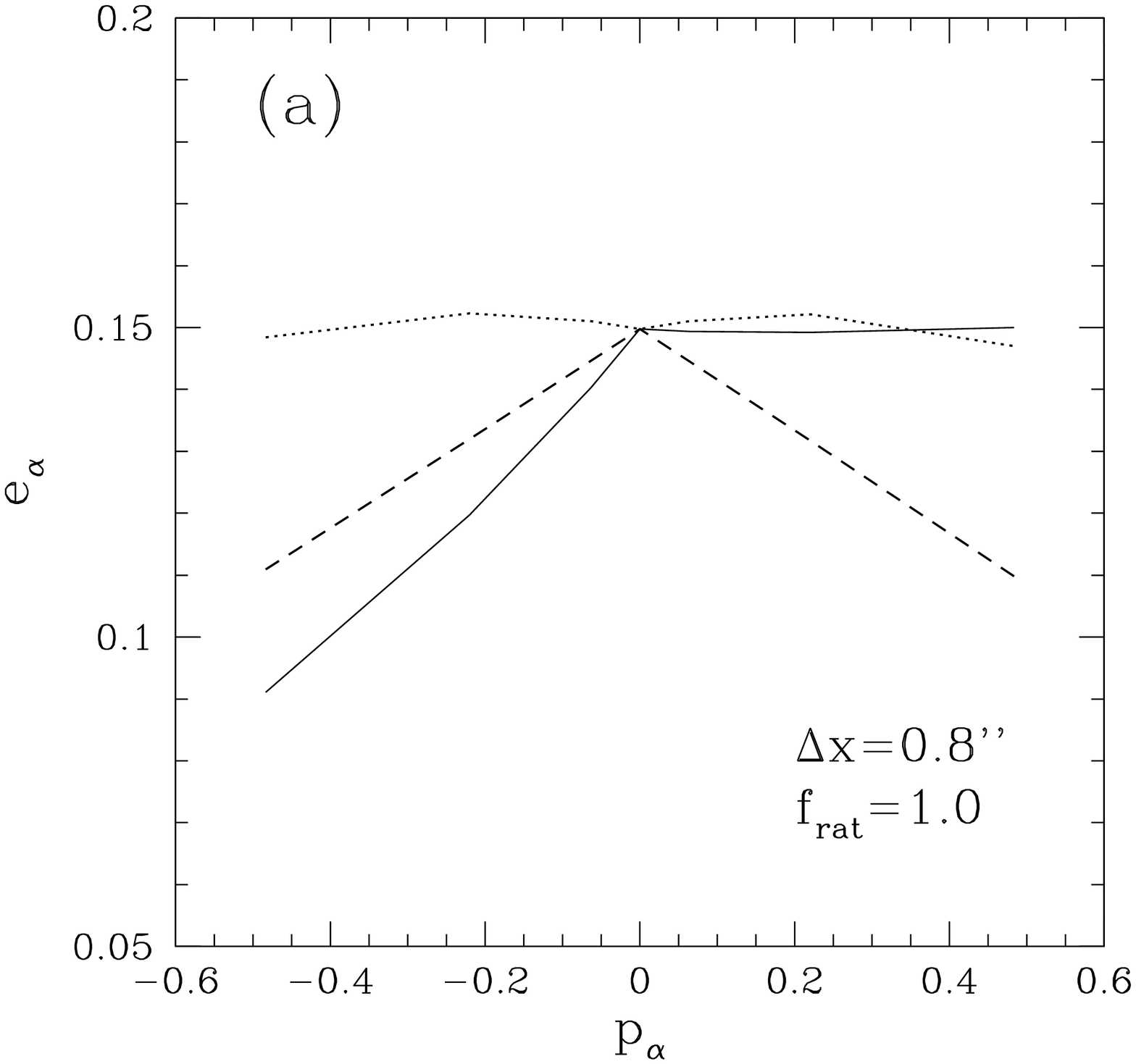}
\epsfxsize=7.5cm
\epsffile{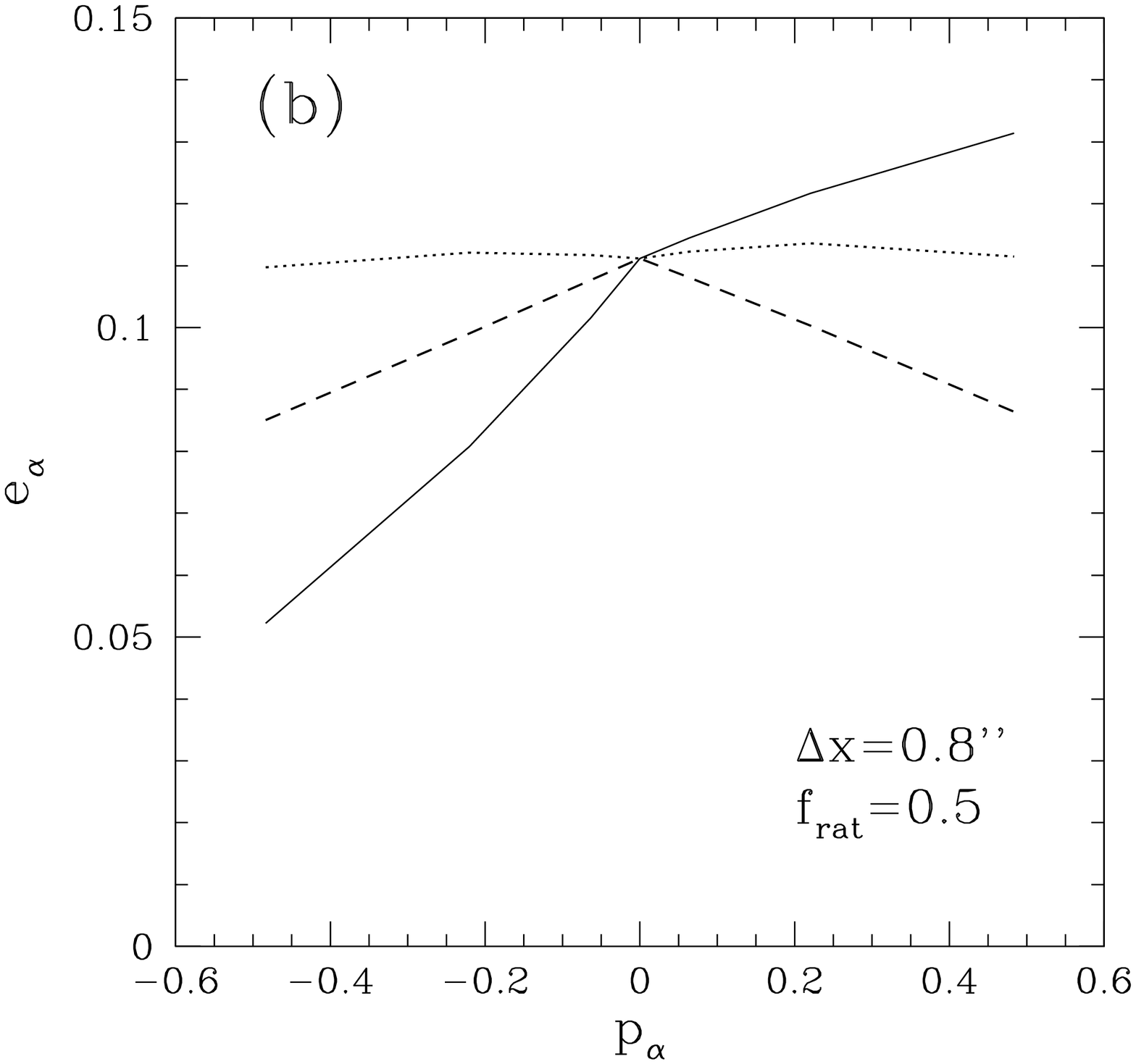}}

\figcaption{\footnotesize (a) Measured polarisation as a function of
PSF anisotropy $p_\alpha$, for two point sources of equal flux,
separated by $0\farcs4$ and a seeing of 1 arcsecond. The thin solid
line indicates the values without any correction for PSF anisotropy.
A single point source would show a linear trend, but because of the
second point source the polarisation is constant for $p_\alpha>0$.
The dashed line corresponds to the polarisation when equation~3 is
used to correct for PSF anisotropy. A clear trend $\propto |p|$ can be
seen (in this case, the PSF anisotropy is given by $p_\alpha$, with
the other component zero). If we use equation~5, with an appropriate
coefficient for the slope, we obtain the dotted line, which is
effectively independent of PSF anisotropy. (b) The same as (a), but
for a second component with a flux ratio of 0.5. The coefficient
to obtain the dotted line is different from the one used in panel~(a).
\label{cormod}}
\end{center}
\vspace{-0.4cm}
\end{figure*}

\section{Improving the correction for PSF anisotropy}

As indicated by Figure~\ref{psfcor} the correction for PSF anisotropy
using equation~3 leaves a systematic residual, roughly proportional to
the polarisation. This correction scheme has been used extensively in
weak lensing applications, and has been tested in great detail. The
difference between the analysis of galaxies and the blends considered
here, is that the shapes of galaxies are well characterized by their
quadrupole moments. In the case of two point sources, higher order
moments contribute to the moments.

In this section we examine how to improve the correction for PSF
anisotropy, in particular we justify the use of equation~5. Unlike the
case for galaxies (Kaiser et al. 1995), this problem is too
complicated to solve analytically. Instead we study the effect of PSF
anisotropy on simulated images of two point sources. We also note that
new methods have been developed in which the images of the objects are
decomposed into a series of localized basis functions. For instance,
Bernstein \& Jarvis (2002) use Laguerre expansions, whereas Refregier
(2003) adopted weighted Hermite polynomials. The advantage of these
methods might be that they can quantify higher order moments of the images.
Nevertheless, as we will show below (and in \S4), the empirical extension
of the Kaiser et al. (1995) method is adequate for the results presented
here.

We create well oversampled images of two point sources, and convolve
these with a Moffat function, with a width given by the required
seeing. These images are then convolved with a ``line'', which
simulates the effect of PSF anisotropy. Examples for two
configurations are indicated by the thin solid lines in
Figure~\ref{cormod}.  The results presented in this Figure are for a
case where the PSF anisotropy is given by $p_\alpha$ alone, with the
other component set to zero. A single point source would show a
linear trend with $p_\alpha$, but because of the second point source
the slope changes when $p_\alpha$ changes sign.

The next step is to correct these polarisations for PSF anisotropy.
If we use equation~3, we obtain the dashed lines in
Figure~\ref{cormod}. In both cases we see a clear residual $\propto
|p_\alpha|$. More general simulations, with both components of the PSF
anisotropy non-zero, indicate that the slope is actually $\propto |p|$,
and not just $|p_\alpha|$. These results have led us to consider
an additional term in the correction $\propto \sqrt{p_1^2+p_2^2}$,
leading to equation~5.

Based on a large set of simulations, we found that the slope of the
trend is proportional to $e_\alpha$. This is not completely
surprising, because the amplitude of the polarisation is a measure of
the importance of the second point source, and consequently a measure
of the relevance of higher order moments. Hence the additional term in
the correction for PSF anisotropy is given by $\gamma e_\alpha |p|$.

We examined what value for $\gamma$ yields the best correction. The
results indicate that $\gamma$ depends on the configuration, in
particular on the flux ratio. For instance, in Figure~\ref{cormod}a we
used $\gamma=0.7$ and in Fig.~\ref{cormod}b we obtained the best
result for $\gamma=0.6$ to obtain the improved corrections, indicated
by the dotted lines. Overall, the range in $\gamma$ appears to be
fairly small, although its value is difficult to determine if the
polarisations are small (i.e., when the additional correction is
small).

We note that the conclusions listed above are based on our empirical
study of the simulated images. using equation~5, the correction for
PSF anisotropy works well for most of the situations (as is the case
for the ones presented in Fig.~\ref{cormod}). In some extreme cases,
however, with large values for the polarisation and PSF anisotropy,
the correction leaves significant residuals (relative errors as large
as 10\%).

\end{document}